# JOINT END-TO-END IMAGE COMPRESSION AND DENOISING: LEVERAGING CONTRASTIVE LEARNING AND MULTI-SCALE SELF-ONNS

*Yuxin Xie[1], Li Yu[2], Farhad Pakdaman[3], and Moncef Gabbouj[3]*

[1]School of Software, [2]School of Computer Science, Nanjing University of Information Science and Technology, China

[3]Faculty of Information Technology and Communication Sciences, Tampere University, Finland

## ABSTRACT

Noisy images are a challenge to image compression algorithms due to the inherent difficulty of compressing noise. As noise cannot easily be discerned from image details, such as high-frequency signals, its presence leads to extra bits needed for compression. Since the emerging learned image compression paradigm enables end-to-end optimization of codecs, recent efforts were made to integrate denoising into the compression model, relying on clean image features to guide denoising. However, these methods exhibit suboptimal performance under high noise levels, lacking the capability to generalize across diverse noise types. In this paper, we propose a novel method integrating a multi-scale denoiser comprising of Self Organizing Operational Neural Networks, for joint image compression and denoising. We employ contrastive learning to boost the network ability to differentiate noise from high frequency signal components, by emphasizing the correlation between noisy and clean counterparts. Experimental results demonstrate the effectiveness of the proposed method both in rate-distortion performance, and codec speed, outperforming the current state-of-the-art.

## 1. INTRODUCTION

Existing image denoising solutions consist of a sequential pipeline where image denoising is performed after the compression and decompression. This has led to independent development and advancement of image compression [1][2][3] and denoising [4][5] techniques. This sequential denoising approach faces two main problems. First, noise is irregular in nature and difficult to compress. Hence, compression algorithms often waste extra bits to store and communicate noise, which reduces bitrate efficiency and visual quality [6]. Second, adding a denoising operation to the decoder side increases the decoding complexity, and time and energy consumption [7]. These problems make this approach challenging to meet the contemporary demands for image processing.

**The emerging Learned Image Compression** approach [1][2] replaces the traditional compression pipeline with a learnable structure, often trained end-to-end. Although these methods are complex and still not ready to take over the traditional standards [8], they provide interesting new opportunities. One such opportunity is that the end-to-end training process enables optimizing the codec for any criteria. This potential has recently been exploited to optimize the codec for image processing tasks.

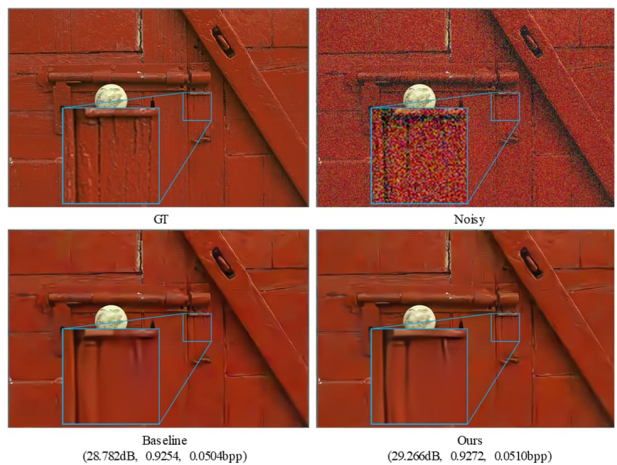

**Fig. 1**. Qualitative comparison of results on the Kodak dataset where the noise level is 4.

The JPEG AI standard [9], which is under development now, has also mandated the ability to process images in compressed domain.

**Compressed domain denoising** directly enhances image quality within the compressed domain while significantly reducing storage and computational demands. Recent methods [7] have integrated convolutional layers into the decoder to eliminate noise from the latent features during decoding. However, this approach still suffers from the encoder allocating bits to store the noise. Moreover, basic convolutional layers are often inefficient in balancing noise removal with the preservation of image details when handling images with various noise levels. Accordingly, a novel two-branch, weight-sharing architecture with plug-in feature denoisers [6] has been proposed, which uses clean features to guide the denoising of noisy features. However, its performance is suboptimal under high noise levels. A similar approach for remote sensing image processing is proposed in [10].

In this paper, we propose a novel end-to-end network that integrates contrastive learning with multi-scale Self Organized Operational Neural Networks (Self-ONNs), optimized for joint image compression and denoising. Inspired by [6], we employ a guidance branch to supervise the denoising at the encoder. Moreover, we propose to use a contrastive loss, which reduces the distance between similar features while increasing the distance with dissimilar ones. Emphasizing the similarity between clean images and their noisy counterparts boosts the model's ability to distinguish between noise and image details. Self-ONNs [11] replace the linear convolution kernel with generative kernels, which find the optimal operator for each neuron through Taylor series expansion. These methods have shown great performance in solving image-to-image translation tasks [12][13], and hence are employed in the denoiser to improve denoising in high noise levels and in presence of detailed texture. We use a multi-scale denoiser structure which not only expands the receptive field but also provides more abundant contextual information. Convolutional Block Attention Modules (CBAM) [14] are used in the denoiser to guide the operations towards important features, and enhance computational efficiency. The proposed method outperforms the state-of-the-art especially at high noise levels, while also offering faster encoding and decoding. Figure 1 depicts an example, showcasing the visual performance of the proposed method. The contributions of this paper are summarized as follows:

- We propose a joint compression-denoising method, which uses contrastive learning and guiding loss, helping the model to better discern noise from details at encoding stage.
- We propose a denoiser module using multi-scale Self-ONN blocks and CBAM attention, which leverages the heterogeneity and enhanced non-linearity of operations to effectively handle images with various noise levels.
- Extensive experiments demonstrate that the proposed method outperforms the state-of-the-art, in rate-distortion performance, especially in dealing with high-level noise, while being faster.

## 2. RELATED WORK

### 2.1. Joint Image Compression and Denoising

While existing imaging solutions perform compression and denoising sequentially, recent proposals suggest handling the two concurrently, aiming to save storage space while enhancing image quality. Testolina et al. [7] proposed integrating convolutional layers into the decoder of a compression model to enable denoising capabilities. Alvar et al. [15] combined denoising and compression, improving bitrate efficiency via latent-space scalability. Xie et al. [6] introduced a two-branch, weight-sharing architecture, intended for noise-aware image compression. The guidance branch uses guiding features to guide the denoised features in the denoising branch throughout the compression process. Nonetheless, the performance of these methods remains unsatisfactory when dealing with images of high noise levels.

### 2.2 Contrastive Learning for Low-level Vision Tasks

Contrastive learning has significantly advanced low-level vision tasks, such as image super-resolution and restoration. Contrastive learning uses a contrastive loss in latent space to enhance the consistency between different augmentations of the same sample. Its aim is to pull the anchor closer to positive samples and push it away from negative samples, thereby facilitating discriminative feature learning. Dual Contrastive Derain-GAN (DCDGAN) [16], an innovative adversarial framework, used dual contrastive learning for effective rain removal and image restoration. Wu et al. [17] proposed a contrastive learning framework specifically designed for single-image super-resolution. Although contrastive learning is effective in many low-level vision tasks, its application in image denoising remains comparatively less explored. Cui et al. [18] introduced Denoising Contrastive Regularization (DCR) to utilize information from noisy and clean images, aligning denoised images closer to clean ones in feature space. Contrastive learning, through the construction and analysis of positive and negative sample pairs, enables the model to align denoised images closer to clean images in feature space.

### 2.3 Self-Organized Operational Neural Networks

To overcome the limitations of CNNs, particularly their localized processing and linear transformations, a recent shift of focus has been towards light and efficient models. Self-ONN [11] offers significant improvements with its enhanced non-linearity and heterogeneous processing. By leveraging Taylor series expansion, it finds the best non-linear mappings at each neuron. Due to the heterogeneity of Self-ONN, it enables individual neurons within each layer to perform distinct transformations, a significant departure from the uniform processing in traditional CNNs. In [12], Self-ONNs were employed in image restoration while demonstrating superior generalization and performance. Malik et al. [13] utilized Self-ONNs for real-world blind image denoising, demonstrating their superior performance over traditional deep CNNs like DnCNN [19]. The promising results in these relevant application domains motivates the use Self-ONNs in the denoiser module of this paper.

## 3. PROPOSED METHOD

This section details the proposed joint image compression and denoising method, which combines contrastive learning with a multi-scale Self-ONN denoiser, as illustrated in Figure 2. In Section 3.1, we detail the three-branch structure of our model, which facilitates denoising in encoder. Then we introduce the proposed denoiser module in Section 3.2. Finally, in Section 3.3, we describe how the denoised features are compressed with the learned image compression model, and formulate the final optimization problem.

### 3.1 Three-branch Structure

**Guidance Branch.** Learned compression methods transform the input image into latent features, which is then compressed with entropy coding. We use a guidance branch to guide the encoder transformation to learn denoised image features. The clean image $x$ is input into encoders $ga_0$ and $ga_1$, extracting guiding features $y_0^{gt}$ and $y_1^{gt}$ for guiding the denoising process of the noisy synthetic image $\tilde{x}$.

$$y_0^{gt} = ga_0(x) \qquad (1)$$

$$y_1^{gt} = ga_1(y_0^{gt}) \qquad (2)$$

**Denoising Branch.** The noisy image $\tilde{x}$ is first processed by encoder $ga_0$ and denoiser $d_0$ to obtain feature $y_0$. It is then further processed by $ga_1$ and $d_1$ to achieve $y_1$. Guiding features $y_0^{gt}$ and $y_1^{gt}$ supervise the denoising, helping to eliminate noise effectively. This process is formulated as follows:

$$y_0 = ga_0(\tilde{x}) + d_0(ga_0(\tilde{x})) \qquad (3)$$

$$y_1 = ga_1(y_0) + d_1(ga_0(y_0)) \qquad (4)$$

The two-level guidance loss $G$ aims to minimize the $L_1$ distance between the guiding features and the denoising features:

$$G = \left\| y_0 - y_0^{gt} \right\|_1 + \left\| y_1 - y_1^{gt} \right\|_1 \qquad (5)$$

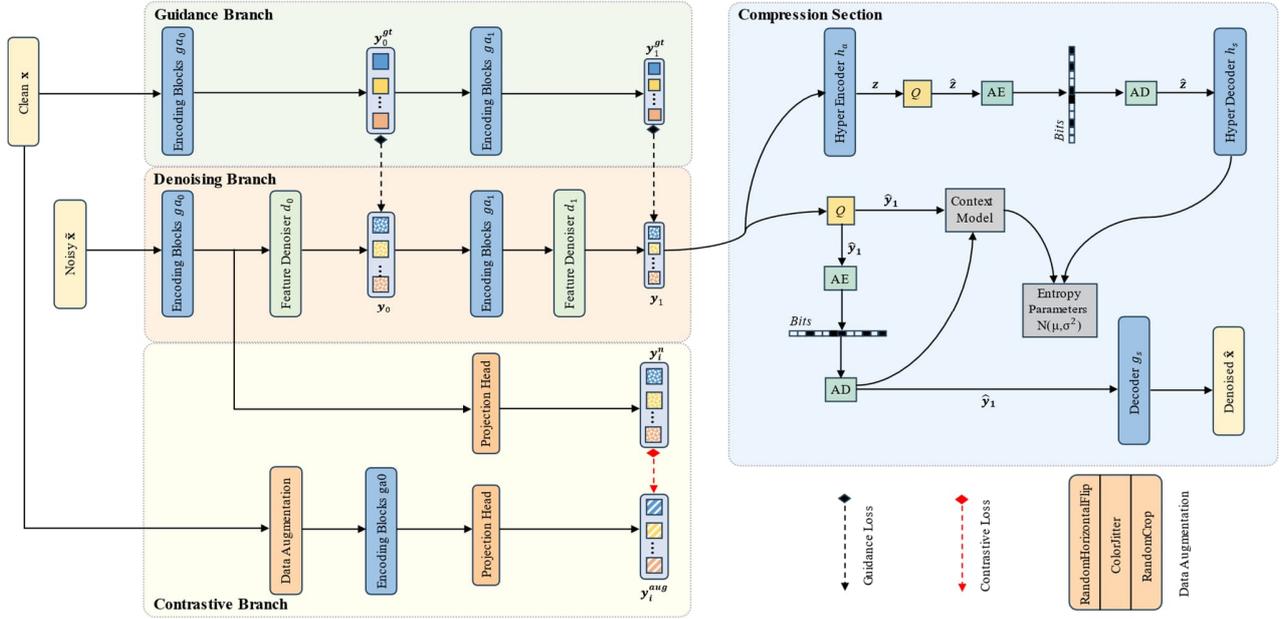

**Fig. 2**. Overview of the proposed method, with a three-branch structure: the guidance branch supervises the denoising process, and the contrastive branch aids the model to discriminate noise from details. Denoised features are compressed with learned compression.

**Contrastive Branch.** By introducing a learnable nonlinear transformation layer [20] between representation and contrastive loss, contrastive learning enhances the training performance. To construct positive and negative sample pairs, two different augmentation methods are applied to the same set of images. Synthesized noisy images from the denoising branch are used as an enhancement to the clean image. A series of data augmentation techniques, such as flip, jitter, and crop, are applied to the same clean image as another form of enhanced sample. These augmentation techniques alter the visual characteristics of the image without introducing extra noise, maintaining the basic content and structure of the image. These enhanced samples form positive sample pairs, emphasizing the similarity between the noisy image and its corresponding clean image. Negative sample pairs are constructed by pairing a noisy image with an enhanced sample that differs in content, specifically any sample other than the noisy image itself. Contrastive learning refines the ability to discern key features of the original image from noise. Representations are mapped to a lower-dimensional space using the projection head, which is an *MLP* module. The model is trained to enhance the similarity between positive sample pairs $y_i^n$ and $y_i^{aug}$, and decrease the similarity between negative sample pairs $y_i^n$ and $y_j^{aug}$. Hence, our method discerns the core content of clean images even with high noise levels, facilitating effective noise reduction. We choose cosine similarity as the similarity calculation function. The loss of contrastive learning can be represented as:

$$L_{CL} = -log \frac{\exp{(y_i^n \cdot y_i^{aug}/\tau)}}{\sum_{j=1}^{2N} \mathbb{1}_{[j \neq i]} exp(y_i^n \cdot y_j^{aug}/\tau)} \quad (6)$$

where $\mathbb{1}_{[j \neq i]}$ is an indicator function that is calculated to be 1 if and only if $j \neq i$, and $\tau$ is a temperature parameter, which determines the focus on hard negative samples, those with higher similarity to positive samples.

### 3.2 Multi-Scale Denoiser using Self-ONN and CBAMs

Our proposed denoiser employs a multi-scale strategy consisting of Self-ONN layers, and CBAM attention mechanism [14], to eliminate residual noise, effectively handling images with various noise levels. Figure 3 depicts the denoiser structure for $d_0$ and $d_1$.

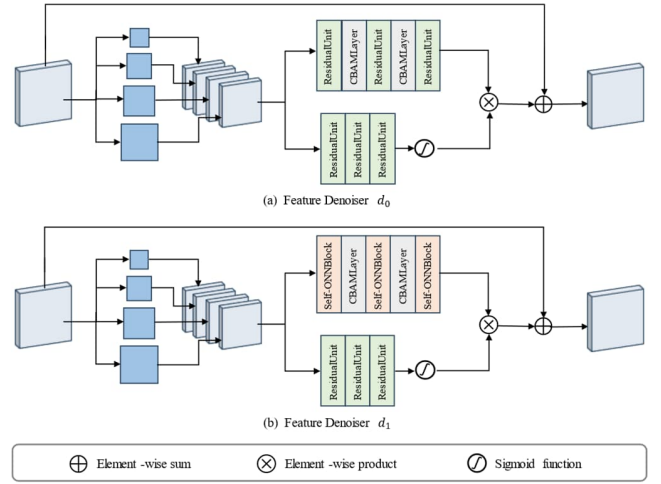

**Fig. 3**. Multi-Scale Self-ONN Denoiser. Our denoiser employs a cascading structure: $d_0$ and $d_1$, integrating a multi-scale strategy.

**Multi-scale strategy.** To overcome the local connectivity of CNNs, as illustrated in Figure 3, we progressively expand the receptive field using convolutional layers with varying dilation rates. Each layer has a different receptive field size, ranging from 3×3 to 9×9, enabling the capture of image information from local details to global features. Taking computational complexity into account, we adjust the channel numbers in each convolutional layer. The output channels are reduced to one-fourth of the input channels and are then merged to align with the original input channel count [4].

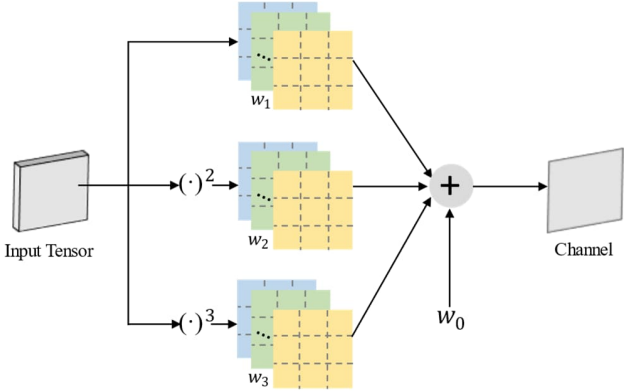

**Fig. 4**. A generative neuron for $Q = 3$. The parameters of each generative neuron, $w_1$, $w_2$ and $w_3$, represent convolutional kernels for the input channels, while $w_0$ serves as the bias term.

**Self-ONN modules.** Self-ONNs use layers of generative neurons to generates optimal operators at each neuron using the Taylor series expansion of $f(x)$:

$$f(x) = \sum_{n=0}^{\infty} \frac{f^{(n)}(a)}{n!}(x-a)^n \quad (7)$$

The $Q^{th}$ order Taylor series truncation, namely the Taylor polynomial, is expressed as follows:

$$f(x)^{(Q,a)} = \sum_{n=0}^{Q} \frac{f^{(n)}(a)}{n!} x^n \quad (8)$$

Where $Q$ is the hyperparameter controlling the degree of Taylor approximation, expanded around $a$. For an input tensor with $c$ channels, the parameters $w_n$, for $n = 1,...,Q$, correspond to $Q$ banks of $c$-channel convolutional kernels, and $w_0$ acts as a bias term. When $a = 0$ and $Q = 1$, the generative neuron simplifies into a conventional convolutional neuron. These parameters can be trained using backpropagation. Figure 4 visualizes a generative neuron with 3×3 kernels, $a = 0$ and $Q = 3$.

We employ a two-stage feature denoiser, $d_0$ and $d_1$, as shown in Figure 3. We introduce the attention mechanism CBAM [14] into the denoiser, focusing on crucial image details. The preliminary denoising is conducted using feature denoiser $d_0$, removing noticeable noise from the input image. Within the feature denoiser $d_1$, Self-ONNs are employed in place of traditional CNNs. The Self-ONN, with its heterogeneity and non-linear capabilities, enables $d_1$ to effectively eliminate residual noise and handle images with varying noise levels.

### 3.3 Learned Compression Model

The denoised features are compressed with the learned compression model in [1]. The latent features $y_1$ are quantized into $\hat{y}_1$. In training, this is replaced by an additive uniform noise $U(-0.5, 0.5)$, to address the nondifferentiable quantization [21]. To capture the spatial dependencies in $y_1$, a hyperprior analysis $h_a$ extracts side information $z$, which is then quantized into $\hat{z}$. The context model $C$, along with the output of the hyper-decoder $h_s$, are combined to predict the distribution parameters used for Arithmetic Encoding (AE) and Decoding (AD) of the latent features $\hat{y}_1$. Learned image compression model aims to minimize the expected length of the bitstream and the distortion of the decoded image $\hat{x}$. The rate-distortion loss function is defined as:

$$\mathcal{L}_{rd} = \mathcal{R}(\hat{y}_1) + \mathcal{R}(\hat{z}) + \lambda_d \mathcal{D}(x, \hat{x}) \quad (9)$$

where $\lambda_d$ controls the rate-distortion tradeoff. Different values of $\lambda_d$ correspond to different bitrates. While $\mathcal{R}$ is approximated as the entropy of information, $D$ is measured as the difference between the output and the input image, which is $MSE(x, \hat{x})$ is this work.

**The overall loss** in this work is formed by combining the rate-distortion loss (given as (9)), the guidance loss (given as (5)), and the contrastive loss (given as (6)). This balances between the compression efficiency and the denoising capability, and guides the compression towards denoised image features. The overall loss function is:

$$\mathcal{L} = \mathcal{R}(\hat{y}_1) + \mathcal{R}(\hat{z}) + \lambda_d \mathcal{D}(x, \hat{x}) + \lambda_g G + \lambda_c \mathcal{L}_{CL} \quad (10)$$

where $\lambda_g$ is the weight of the guidance loss, and $\lambda_c$ controls the weight of the contrastive learning. These two are set to 3 and 1.5 in experiments, respectively.

## 4. EXPERIMENTS

### 4.1 Experiment Setup

**Implementation Details.** The proposed method is implemented with PyTorch, using the CompressAI library [3]. Our models are fine-tuned on the pre-trained cheng2020 [1], on an NVIDIA GeForce RTX 4090 Ti. For fair comparisons, we use the same hyperparameters as the baseline [6] for training over six quality points. The channel numbers for low bitrate models (q1, q2, q3) were set to 128, and for high bitrate models (q4, q5, q6) to 192 [1].

**Synthetic Datasets.** To approximate real-world noise, sRGB images are converted to raw RGB format, simulating the two main types of noise in real camera sensors: readout noise parameter $\sigma_r$ and shot noise parameter $\sigma_s$. During training, $\sigma_r$ and $\sigma_s$ were uniformly sampled from the ranges $[10^{-3}, 10^{-1.5}]$ and $[10^{-4}, 10^{-2}]$, respectively. In validation and testing, four predetermined pairs of parameters $(\sigma_r, \sigma_s)$ from the official test set [5] were used, which are named hereafter noise levels 1 to 4. We utilize the Flicker 2W dataset [22] for training and validation. All trained models are evaluated on the Kodak [23] and CLIC [24] datasets, which are commonly used for image processing methods.

**Evaluation Metrics.** To assess the rate-distortion performance, we calculated the Bits Per Pixel (bpp) and distortion metrics, including the Peak Signal-to-Noise Ratio (PSNR) and the Multiscale Structural Similarity Index (MS-SSIM).

### 4.2 Denoising on Synthetic Noisy Images

**Qualitative Comparison.** Visual comparisons are used to evaluate the effectiveness of the proposed method, as illustrated in Figures 1 and 5. We selected some challenging examples from Kodak and CLIC datasets at a noise level of 4, and with similar bpps. It is observed that at high noise levels, distinguishing features such as the gaps in wood boards, lines on roof eaves, and textures on backpacks become challenging to discern from the noise for the baseline method. The images provide clear evidence that our results maintain superior details and exhibit sharper textures, indicating a notable improvement over the baseline [6] method.

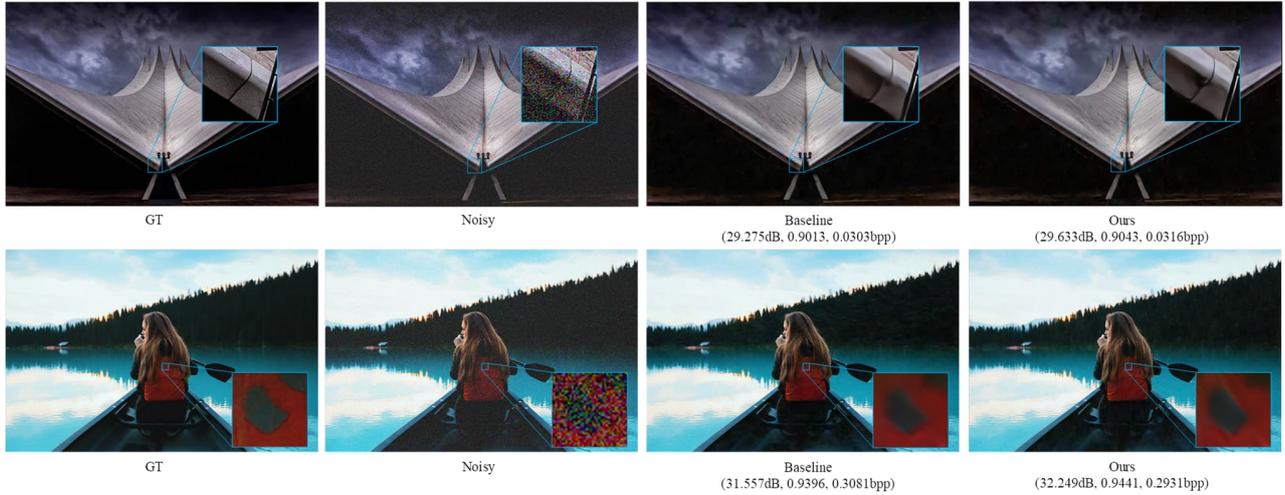

**Fig. 5**. Qualitative comparison of results on the CLIC dataset at noise level 4. From left to right, the sequence is the ground truth, the noisy image, the results of baseline [6], and ours.

**Table 1**. The BD-Rate PSNR and BD-Rate MS-SSIM of the proposed method compared to Baseline [6].

| Noise level | BD-Rate PSNR | | BD-Rate MS-SSIM | |
|---|---|---|---|---|
| | Kodak | CLIC | Kodak | CLIC |
| level 1 | -0.45% | -0.37% | -0.19% | -0.45% |
| level 2 | -0.74% | -0.55% | -0.62% | -0.85% |
| level 3 | -1.35% | -1.25% | -1.27% | -1.93% |
| level 4 | -23.80% | -27.85% | -15.40% | -72.19% |

**Quantitative Comparison.** Rate-Distortion (RD) curves, derived from the experimental results, are used to evaluate the performance of competing methods. As illustrated in Figure 6, we compare the proposed method against the baseline [6], and a sequential denoise-then-compress method, for which, denoising model DeamNet [25] is used followed by compression using Cheng2020 [1]. It is observed that in noise level 4, the proposed method consistently outperforms the baseline, in all bpps. Similar observation is made based on PSNR and MS-SSIM. For the sequential method, it can be observed that the performance significantly drops with the increase of bitrate. This is mainly due to the limitation of denoiser module trained individually, which cannot generalize on all noise levels and compression ratios, and consistent with findings in [6].

To quantify the RD performance, Table 1 summarizes the Bjøntegaard Delta Rate (BD-Rate) based on PSNR and BD-Rate based on MS-SSIM of our method, compared to baseline on the Kodak and CLIC datasets. The table confirms the observations from Figure 6 with large BD-Rate savings for noise level 4. For noise level 3, the gains are much lower, but still as high as 1.93%. For noise levels 2 and 1, this value decreases to lower than 1%. The lower gain in lower noise levels is justified by the fact that (1) in lower noise levels, noise contributes less to the overall bitrate, and hence, limited gain is achieved by removing noise. (2) The baseline [6] method performs well for low noise levels and manages to keep the quality while compressing well. Hence, its performance is closer to the proposed method.

**Computational Complexity.** We compare the complexity of our method with the baseline method [6] on Kodak and CLIC datasets, as summarized in Tables 2 and 3. All tests are repeated 3 times, and the average encoding and decoding times are reported. We report the average times and the number of model parameters under compression qualities q2 and q4, and noise levels 3 and 4. The tables show that the proposed method, despite having a slightly larger number of parameters, is 3% to 6.48% faster in encoding and decoding compared to the baseline. The main reason for the faster operation is the efficiency of the proposed denoiser module consisting of Self-ONN and CBAM attention, which is able to focus on more important regions. Moreover, it is worth noting that as demonstrated by [6] and [7], joint compression-denoising methods are already much faster than the traditional sequential methods, which is further improved by the proposed method.

**Table 2**. Comparing computational complexity with baseline [6] on quality q2 and q4, on the Kodak dataset.

| Method | Enc /Dec(s) | | Param. |
|---|---|---|---|
| | Level 3 | Level 4 | |
| baseline-q2 | 12.53/25.92 | 12.52/25.93 | 1250k |
| baseline-q4 | 23.51/38.12 | 23.71/38.32 | 2811k |
| ours-q2 | 12.13/24.51 (-3.17%/-5.44%) | 12.14/24.49 (-3.02%/-5.55%) | 1315k |
| ours-q4 | 22.76/36.55 (-3.17%/-4.12%) | 22.37/36.51 (-5.66%/-4.72%) | 2954k |

**Table 3**. Comparing computational complexity with baseline [6] on quality q2 and q4, on CLIC dataset.

| Method | Enc /Dec(s) | | Param. |
|---|---|---|---|
| | Level 3 | Level 4 | |
| baseline-q2 | 74.59/151.27 | 74.68/151.77 | 1250k |
| baseline-q4 | 145.40/235.44 | 145.28/235.67 | 2811k |
| ours-q2 | 72.28/144.03 (-3.10%/-4.79%) | 72.39/142.92 (-3.06%/-5.83%) | 1315k |
| ours-q4 | 140.69/220.35 (-3.24%/-6.41%) | 139.44/220.40 (-4.02%/-6.48%) | 2954k |

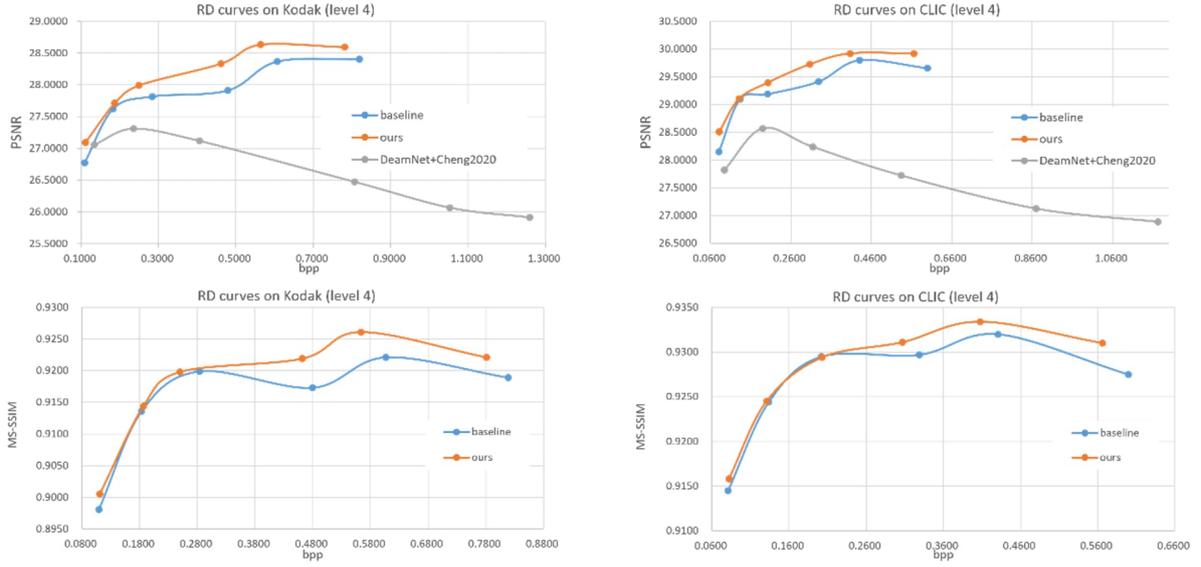

**Fig. 6**. RD curves of the proposed, baseline[6], and a sequential method on Kodak and CLIC datasets at noise level 4.

### 4.3 Ablation study

Ablation studies are conducted on Kodak and CLIC datasets at noise level 4 and quality q6. The results are summarized in Tables 4 and 5. We evaluate different versions of the proposed method, with and without the contrastive learning, Self-ONNs, and the multi-scale denoising structure. The tables reveal a performance drop when contrastive learning is off, thereby demonstrating its effectiveness. Without contrastive learning the model faces challenges in accurately differentiating between noise and details in images.

Ablations on Self-ONN are done by replacing them with CNN layers similar to baseline [6]. It is observed that using Self-ONN improves the performance compared to CNN. Similar observation is made for the multi-scale denoising structure. Each of these modules help the proposed method to model the noise, either through efficient non-linear mapping or a larger receptive field.

**Table 4.** Ablation study on different components. Results on Kodak dataset at Noise Level 4.

| Contrastive learning | Self-ONN | Multi-scale strategy | PSNR | bpp |
|---|---|---|---|---|
| | | | 28.4045 | 0.8189 |
| ✓ | | | 28.4300 | 0.7935 |
| | ✓ | | 28.3270 | 0.8372 |
| ✓ | ✓ | | 28.5306 | 0.7946 |
| ✓ | ✓ | ✓ | 28.5949 | 0.7812 |

**Table 5.** Ablation study on different components. Results on CLIC dataset at Noise Level 4.

| Contrastive learning | Self-ONN | Multi-scale strategy | PSNR | bpp |
|---|---|---|---|---|
| | | | 29.6535 | 0.6000 |
| ✓ | | | 29.6725 | 0.5880 |
| | ✓ | | 29.5828 | 0.6046 |
| ✓ | ✓ | | 29.7765 | 0.5814 |
| ✓ | ✓ | ✓ | 29.9215 | 0.5662 |

### 5. CONCLUSION

In this paper, we introduced a novel method for joint image compression and denoising. The proposed method integrates enhanced multi-scale Self-ONN-based denoisers into image compression, and boosts the training with a guidance loss and contrastive learning. Extensive experiments on two datasets confirm the effectiveness of the proposed method. It was observed that contrastive learning addresses the challenge of differentiating noise from details in noisy images, especially in higher noise levels. It was demonstrated that using the compact Self-ONN together with CBAM attention improves the denoising performance while lowering the computational complexity. Results revealed that the proposed method significantly outperforms the state-of-the-art at high noise levels, by restoring richer textures. In lower noise levels, the proposed achieves a higher performance, but not a major improvement due to limited noise effect on compression.

**ACKNOWLEGMENT** This project has received funding from the European Union's Horizon 2020 research and innovation programme under the Marie Skłodowska-Curie grant agreement No [101022466], and from the NSF-Business Finland Center for Big Learning (CBL), Advanced Machine Learning for Industrial Applications (AMaLIA) under Grant 97/31/2023.

### 6. REFERENCES

[1] Zhengxue Cheng, Heming Sun, Masaru Takeuchi, and Jiro Katto, "Learned image compression with discretized gaussian mixture likelihoods and attention modules," in Proceedings of the IEEE/CVF conference on computer vision and pattern recognition, 2020, pp.7939–7948.

[2] Jun-Hyuk Kim, Byeongho Heo, and Jong-Seok Lee, "Joint global and local hierarchical priors for learned image compression," in Proceedings of the IEEE/CVF Conference

on Computer Vision and Pattern Recognition, 2022, pp. 5992–6
[3] Jean Bégaint, Fabien Racapé, Simon Feltman, and Akshay Pushparaja, "Compressai: a pytorch library and evaluation platform for end-to-end compression research," arXiv preprint arXiv:2011.03029, 2020.
[4] Yuanbiao Gou, Peng Hu, Jiancheng Lv, Joey Tianyi Zhou, and Xi Peng, "Multi-scale adaptive network for single image denoising," Advances in Neural Information Processing Systems, vol. 35, pp. 14099–14112, 2022.
[5] Ben Mildenhall, Jonathan T Barron, Jiawen Chen, Dillon Sharlet, Ren Ng, and Robert Carroll, "Burst denoising with kernel prediction networks," in Proceedings of the IEEE conference on computer vision and pattern recognition, 2018, pp. 2502–2510.
[6] Ka Leong Cheng, Yueqi Xie, and Qifeng Chen, "Optimizing image compression via joint learning with denoising," in European Conference on Computer Vision. Springer, 2022, pp. 56–73.
[7] Michela Testolina, Evgeniy Upenik, and Touradj Ebrahimi, "Towards image denoising in the latent space of learning-based compression," in Applications of Digital Image Processing XLIV. SPIE, 2021, vol. 11842, pp. 412–422.
[8] Farhad Pakdaman and Moncef Gabbouj, "Comprehensive complexity assessment of emerging learned image compression on cpu and gpu," in ICASSP 2023-2023 IEEE International Conference on Acoustics, Speech and Signal Processing (ICASSP). IEEE, 2023, pp. 1–5.
[9] João Ascenso, Elena Alshina, and Touradj Ebrahimi, "The jpeg ai standard: Providing efficient human and machine visual data consumption," Ieee Multimedia, vol. 30, no. 1, pp. 100–111, 2023.
[10] Vinicius Alves de Oliveira, Marie Chabert, Thomas Oberlin, Charly Poulliat, Mickael Bruno, Christophe Latry, Mikael Carlavan, Simon Henrot, Frederic Falzon, and Roberto Camarero, "Satellite image compression and denoising with neural networks," IEEE Geoscience and Remote Sensing Letters, vol. 19, pp. 1–5, 2022.
[11] Serkan Kiranyaz, Junaid Malik, Habib Ben Abdallah, Turker Ince, Alexandros Iosifidis, and Moncef Gabbouj, "Self-organized operational neural networks with generative neurons," Neural Networks, vol. 140, pp. 294–308, 2021.
[12] Junaid Malik, Serkan Kiranyaz, and Moncef Gabbouj, "Self-organized operational neural networks for severe image restoration problems," Neural Networks, vol. 135, pp. 201–211, 2021.
[13] Junaid Malik, Serkan Kiranyaz, Mehmet Yamac, Esin Guldogan, and Moncef Gabbouj, "Convolutional versus self-organized operational neural networks for real-world blind image denoising," arXiv preprint arXiv:2103.03070, 2021.
[14] Yana Luo and Zhongsheng Wang, "An improved resnet algorithm based on cbam," in 2021 International Conference on Computer Network, Electronic and Automation (ICCNEA). IEEE, 2021, pp. 121–12
[15] Saeed Ranjbar Alvar, Mateen Ulhaq, Hyomin Choi, and Ivan V Bajić, "Joint image compression and denoising via latent-space scalability," arXiv preprint arXiv:2205.01874, 2022.
[16] Xiang Chen, Jinshan Pan, Kui Jiang, Yufeng Li, Yufeng Huang, Caihua Kong, Longgang Dai, and Zhentao Fan, "Unpaired deep image deraining using dual contrastive learning," in Proceedings of the IEEE/CVF Conference on Computer Vision and Pattern Recognition, 2022, pp. 2017–2026.
[17] Gang Wu, Junjun Jiang, and Xianming Liu, "A practical contrastive learning framework for single-image super-resolution," IEEE Transactions on Neural Networks and Learning Systems, 2023.
[18] Taoyong Cui and Yuhan Dong, "Contrastive learning for low-light raw denoising," arXiv preprint arXiv:2305.03352, 2023.
[19] Kai Zhang, Wangmeng Zuo, Yunjin Chen, Deyu Meng, and Lei Zhang, "Beyond a gaussian denoiser: Residual learning of deep cnn for image denoising," IEEE transactions on image processing, vol. 26, no. 7, pp. 3142–3155, 2017.
[20] Ting Chen, Simon Kornblith, Mohammad Norouzi, and Geoffrey Hinton, "A simple framework for contrastive learning of visual representations," in International conference on machine learning. PMLR, 2020, pp. 1597–1607.
[21] Johannes Ballé, Valero Laparra, and Eero P Simoncelli, "End-to-end optimized image compression," arXiv preprint arXiv:1611.01704, 2016.
[22] Jiaheng Liu, Guo Lu, Zhihao Hu, and Dong Xu, "A unified end-to-end framework for efficient deep image compression," arXiv preprint arXiv:2002.03370, 2020.
[23] Eastman Kodak Company. 1999. Kodak Lossless True Color Image Suite. http: //r0k.us/graphics/kodak/
[24] George Toderici, Wenzhe Shi, Radu Timofte, Johannes Ballé, Eirikur Agustsson, Nick Johnston, and Fabian Mentzer, "Workshop and challenge on learned image compression," in Proceedings of the IEEE/CVF Conference on Computer Vision and Pattern Recognition, 2020.
[25] Chao Ren, Xiaohai He, Chuncheng Wang, and Zhibo Zhao, "Adaptive consistency prior based deep network for image denoising," in Proceedings of the IEEE/CVF conference on computer vision and pattern recognition, 2021, pp. 8596–8606.